\documentclass[aps, prl, twocolumn, superscriptaddress, showpacs] {revtex4}

\usepackage{epsfig,amsfonts,amssymb,amsmath}

\newcommand{\eqn}[1]{\begin{equation} #1 \end{equation}} % equation environment
\newcommand{\aln}[1]{\begin{align} #1 \end{align}}       % align environment
\newcommand{\mc}{\mathcal}                               % script letters
\newcommand{\mt}[1]{\hbox{\large {\texttt #1}}}          % straight letters
\newcommand{\eq}[1]{(\ref{#1})}                % equation reference
\newcommand{\e}{\ensuremath{\mathrm{e}}}       % straight e
\renewcommand{\d}{\ensuremath{\mathrm{d}}}     % straight d
\newcommand{\wtilde}{\widetilde}               % wide tilde
\newcommand{\wbar}{\overline}                  % wide bar

\begin{document}

%\preprint{}

\title{The role of orbital dynamics in spin relaxation\\
       and weak antilocalization in quantum dots }



\author{Oleg Zaitsev}
\email[E-mail: ]{oleg.zaitsev@physik.uni-regensburg.de}
%\thanks{}
\affiliation{Institut f\"ur Theoretische Physik, Universit\"at Regensburg,
D-93040 Regensburg, Germany}

\author{Diego Frustaglia}
\affiliation{NEST-INFM \& Scuola Normale Superiore, 56126 Pisa, Italy}

\author{Klaus Richter}
\affiliation{Institut f\"ur Theoretische Physik, Universit\"at Regensburg,
D-93040 Regensburg, Germany}


%\date{\today}

\begin{abstract}
We develop a semiclassical theory for spin-dependent quantum transport to
describe weak (anti)localization in quantum dots with spin-orbit coupling. This
allows us to distinguish different types of spin relaxation in systems with
chaotic, regular, and diffusive orbital classical dynamics. We find, in
particular, that for typical Rashba spin-orbit coupling strengths, integrable
ballistic systems can exhibit weak localization, while corresponding chaotic
systems show weak antilocalization. We further calculate the
magnetoconductance and analyze how the weak antilocalization is suppressed
with decreasing quantum dot size and increasing additional in-plane magnetic
field.
\end{abstract}

%\keywords{}

\maketitle

%\section{}
%\subsection{}
%\subsubsection{}

Weak localization (WL) and antilocalization (AL) are classic examples for
quantum interference and spin-orbit (SO) interaction effects on the conductance
in low-dimensional electronic systems \cite{berg2, chak}. Very recently,  
%with
%the increasing interest in spin phenomena in nano-devices, 
particularly weak AL
has been reconsidered in a number of experiments since AL can be employed as a
probe of SO-induced spin dynamics and relaxation phenomena.  Measurements have
been performed both for GaAs- and InAs-based two-dimensional (2D) electron
gases \cite{schier}, as well as for ballistic bismuth \cite{hackens} and GaAs
\cite{zumb} cavities.  While SO scattering in extended disordered systems is
well understood \cite{chak}, the latter experiments  address the timely
question of how quantum confinement of the orbital motion affects spin
relaxation in clean ballistic quantum dots where the elastic mean-free path is
much larger than the system size. Considerable related progress has
also been made theoretically in treating spin relaxation and the
interplay between SO and Zeeman coupling in quantum dots \cite{khae} including
random-matrix theory (RMT) \cite{alei,crem}. However, the RMT results apply
only to chaotic quantum dots and contain geometric parameters
which must be obtained by other means for a given system.

Here we present an alternative, semiclassical theory for the spin-dependent
magnetoconductance of quantum dots, i.e. a semiclassical Landauer formula
including spin, and apply it to describe weak AL in 2D confined systems. This
approach  allows to uncover the inter-relation between orbital dynamics and
spin evolution in a transparent way, and it is rather generally applicable to
quantum dots with different type of classical dynamics, e.g., chaotic and
regular. Remarkably, we find significant qualitative differences in the spin
relaxation times of chaotic, integrable, and open diffusive systems: Spin
relaxation for confined chaotic systems is much slower than for diffusive
motion; moreover, for a number of integrable geometries we even find  a
saturation, i.e., a certain spin polarization is preserved. Furthermore, we
examine the effect of the system size and of an additional in-plane magnetic
field on the resulting AL. 

Our study is based on the semiclassical Landauer formula \cite{bara,rich} that
we generalize to systems with SO and Zeeman interaction. To this end,
we extend techniques for spin semiclassics, recently
developed for the density of states \cite{bolt,plet}, to quantum transport.
We consider a Hamiltonian linear in the spin operator $\hat {\mathbf s}$,
\eqn{
  \hat H = \hat H_0 (\hat {\mathbf q}, \hat {\mathbf p}) + \hbar\, \hat
  {\mathbf s} \cdot \hat {\mathbf C} (\hat {\mathbf q}, \hat {\mathbf p}),
\label{soHam}
}
where $\hat {\mathbf C} (\hat {\mathbf q}, \hat {\mathbf p})$ is a vector
function of the position and momentum operators $\hat {\mathbf q},\, \hat
{\mathbf p}$, which may include an external (inhomogeneous) magnetic field. 
For a large number of systems of interest, and usually in experiments,
$\hbar\, s\,  |\mathbf C (\mathbf q, \mathbf p)| \ll H_0$, 
even if the spin-precession length is of the order of the system size.
Here $s$ is the particle spin, and
the phase-space functions without a hat denote the classical counterparts 
(Wigner-Weyl symbols) of the respective operators.
As a consequence of the above inequality
the influence of spin on the orbital motion can be neglected.
%\footnote{Formally this limit (assumed in
%this paper) is realized by taking $\hbar \to 0$, while keeping
%all other quantities finite which implies that the
%orbital subsystem $H_0$ is in the semiclassical regime, i.e. the typical
%action $\mc S \gg \hbar$.}.
Thus $H_0$ determines the classical trajectories $\gamma = (\mathbf q (t),
\mathbf p (t))$ 
which, in turn, generate an effective time-dependent magnetic field $\mathbf
C_\gamma (t) = \mathbf C (\mathbf q (t), \mathbf p (t))$  acting on spin via
the Hamiltonian $\hat H_\gamma (t) = \hbar\, \hat {\mathbf s} \cdot \mathbf
C_\gamma (t)$. Hence the spin dynamics can be treated
\emph{quantum-mechanically} in terms of a (time-ordered)  propagator~$\hat
K_\gamma (t)\! =\! T  \exp[-i \int_0^t dt^\prime  \hat {\mathbf s} \cdot
\mathbf C_\gamma (t^\prime) ]$. 
%In this manner, a weak SO coupling was
%incorporated into the  Gutzwiller trace formula for the semiclassical density
%of states for $s\!=\!1/2$ \cite{bolt} and later for arbitrary spin 
%\cite{plet}. 

To derive a semiclassical expression for the spin-de\-pen\-dent conductance of
a quantum dot, we start from the Landauer formula in two dimensions relating
the two-terminal conductance $G \!=\! (e^2/h) \mc T$  to its transmission
coefficient  \cite{fish}
\begin{equation}
\label{qutran}
  \mc T = \sum_{n=1}^{N^\prime} \sum_{m=1}^N \sum_{\sigma, \sigma^\prime=-s}^s
  \left| t_{n\sigma^\prime, m\sigma} \right|^2 \, .
\end{equation}
The leads support $N$ and $N^\prime$ open orbital channels $m$ and $n$, 
respectively, and we distinguish $2s+1$ spin polarizations in the leads, 
labeled by $\sigma = -s, \ldots, s$. 
In \eq{qutran}, $t_{n\sigma^\prime, m\sigma}$ is the transition amplitude 
between the incoming channel $|m, \sigma \rangle$ and outgoing channel $|n,
\sigma^\prime \rangle$; a corresponding equation holds for the reflection
coefficient $\mc R$ satisfying the normalization condition $\mc T + \mc R =
(2s+1) N$.

A semiclassical evaluation of the transition amplitudes, starting from
a path-integral representation of the Green function, yields 
\cite{zait2} (see \cite{bara} for the spinless case)
\begin{equation}
  t_{n\sigma^\prime, m\sigma} = \sum_{\gamma (\bar n, \bar m)}(\hat
  K_\gamma)_{\sigma^\prime \sigma}\,  \mc A_\gamma \exp \left( \frac i \hbar
  \mc S_\gamma \right) \, ,
\label{semtnm}
\end{equation}
given as coherent summation over classical paths at fixed energy \cite{chan};
corresponding results hold for the reflection amplitudes in terms of 
back-reflected paths.  The sum runs over classical trajectories 
$\gamma (\bar n\!=\!\pm n, \bar m\!=\! \pm m)$
that enter (exit) the cavity at  ``quantized''
angles $\Theta_{\bar m}$ ($\Theta_{\bar n}$).
For hard-wall boundary conditions, 
$\sin \Theta_{\bar m} = \bar m \pi / k w$ and $\sin
\Theta_{\bar n} = \bar n \pi / k w'$, where $k$ is the wavenumber,
and $w, w'$ are the lead widths. In \eq{semtnm}, 
$\mc S_\gamma $ ($=\! \hbar k L_\gamma$ for billiards) is the action along 
$\gamma$ with time $T_\gamma$ and classical weight
$\mc A_\gamma$ \cite{bara}.
The entire spin effect is contained in the matrix elements 
$(\hat K_\gamma)_{\sigma^\prime \sigma}$ of the spin 
propagator $\hat K_\gamma \equiv \hat K_\gamma (T_\gamma)$
between the initial and final spin states.

Inserting \eq{semtnm} into \eq{qutran} we derive  the
semiclassical Landauer formula for spin-dependent magnetotransport 
(including SO and Zeeman interaction):
\eqn{
  \mc T = 
  \sum_{nm} \sum_{\gamma (\bar n, \bar m)} \sum_{\gamma^\prime
  (\bar n, \bar m)} \mc \!\! {\mc M_{\gamma, \gamma^\prime} } \,
  \mc A_\gamma \mc A^*_{\gamma^\prime}
   e^{(i/\hbar) (\mc S_\gamma - \mc S_{\gamma^\prime} ) } \, .
\label{spinTR}
}
The orbital contribution of each pair of paths is weighted by the 
spin \emph{modulation factor}
\eqn{
  \mc M_{\gamma, \gamma^\prime} = \text{Tr}\, (\hat K_\gamma \hat
  K_{\gamma^\prime}^\dag) \, ,
\label{Modf}
}
where the trace is taken in spin space.

WL and AL effects are obtained after energy average of ${\mc T}(E,{\bf B})$
for ballistic quantum dots (subject to an external arbitrarily directed
magnetic field ${\bf B}$).  The leading contributions after
averaging \eq{spinTR} for a chaotic cavity with time-reversal symmetry,
i.e., ${\bf B}  = 0$, are as follows:

(i) The \emph{classical} part consists of the terms $\gamma^\prime =
\gamma$~\cite{bara}, for which the rapidly varying energy-dependent phase  in
the exponent of \eq{spinTR} disappears. Then the modulation factor is $\mc
M_{\gamma, \gamma} = \text{Tr}\, (\hat K_\gamma \hat K_\gamma^\dag) = 2s + 1$,
independent of SO interaction, and reduces to the trivial spin degeneracy. 

(ii) The \emph{diagonal} quantum correction is defined for the
reflection only. It contains the terms with $n=m$ and $\gamma^\prime =
\gamma^{-1}$, where $\gamma^{-1}$ is the time-reversal of $\gamma$~\cite{bara}.
Again, the orbital phases of the trajectory pair cancel; however, the
modulation factor is $\mc M_{\gamma, \gamma^{-1}} \! = \!
\text{Tr}\, (\hat K_\gamma^{\,2})$. 

(iii)~The \emph{loop} contribution comes from pairs of long orbits that stay
close to each other in  configuration space, thereby have nearly equal actions,
and hence persist upon energy average. One orbit of the pair has a
self-crossing with a small crossing angle, thus forming a loop, while its
partner exhibits an ``anticrossing''. Outside the crossing region the orbits
are located exponentially close to each other: the paths are related by
time reversal  along the loop and coincide along the rest of the
trajectories~\cite{rich,sieb}. We have computed the modulation factor for
$\gamma$ and $\gamma^\prime$ neglecting  the crossing region and found  $\mc
M_{\gamma, \gamma^\prime} = \text{Tr}\, (\hat K_l^{\,2})$,  where $l$ is the
loop segment of $\gamma$ \cite{zait2}. 

For spinless particles in chaotic quantum dots  the three contributions to the
averaged transmission and  reflection yield \cite{rich}, for $N = N^\prime \gg
1$,   (i) $\mc T_{\text{cl}}^{(0)} = \mc R_{\text{cl}}^{(0)} = N/2$,  (ii)
$\delta \mc R_{\text{diag}}^{(0)} = 1/2$, and (iii) $\delta \mc
T_{\text{loop}}^{(0)} = \delta \mc R_{\text{loop}}^{(0)} = -1/4$, in agreement
with RMT.  Here the superscript refers to zero spin and zero magnetic field.

In the following, we consider the case of an additional uniform,  arbitrarily
directed magnetic field $\mathbf B$ in the presence of  SO interaction. Besides
the Zeeman interaction, the field component $B_z$ perpendicular to the cavity
generates an additional Aharonov-Bohm (AB) phase factor  $\varphi \!=\! \exp( i
4 \pi \mt A_\gamma B_z/ \Phi_0)$ in the  diagonal and loop terms in Eq.\
\eq{spinTR}.  Here,  $\mt A_\gamma \equiv {\int \mathbf A \cdot \d \mathbf l} /
B_z$  is the effective enclosed area for the diagonal  (loop) contributions
accumulated along the orbits $\gamma$ (loops $l$) neglecting bending of the
orbits, and $\Phi_0 = hc/e$ is the flux quantum.

For broken time-reversal symmetry, e.g., by the perpendicular $\bf B$-field, 
$\mc M_{\gamma, \gamma^\prime}$ should be calculated directly from \eq{Modf}.
We then introduce a generalized modulation factor, $\mc M_\varphi \equiv \mc
M_{\gamma,\gamma^\prime}\varphi$, which  is distributed according to a function
$P (\mc M_\varphi; L, \mathbf B)$,  where $L$ is the trajectory (loop) length
in the diagonal (loop) contribution, and the $\mathbf B$-dependence includes
both the AB phase and the Zeeman interaction. Thus we can define a spin
modulation factor  $  \wbar {\mc M_\varphi} (L; \mathbf B) $
averaged over an ensemble of trajectories with fixed length $L$. 

In chaotic systems, the length distribution is given by $\exp\, (-L/
L_{\text{esc}})$ \cite{bara}, if the escape length $L_{\text{esc}} = \pi \mt
A_c /(w + w^\prime)$, the average length the particle traverses before leaving
the cavity of area $\mt A_c$, is much larger than $L_b$, the average distance
between two consecutive bounces at the boundaries.  
It can be shown \cite{rich,zait2} that the
relevant distribution of loop lengths is determined by the same exponent.  As
in  the case without spin \cite{bara,rich}, the product of the AB phase and
spin modulation factors in Eq.\ \eq{spinTR} can be  eventually substituted by
its average $\langle \wbar {\mc M_\varphi} \rangle_L$ over $L$ and pulled out
of the sum. Thereby we obtain, as relative quantum  corrections for the spin-
and $\bf B$-field-dependent transmission and reflection,
\aln{
  &\delta \mc R_{\text{diag}}/ \delta \mc R_{\text{diag}}^{(0)} = \delta \mc
  R_{\text{loop}}/ \delta \mc R_{\text{loop}}^{(0)} =  \delta \mc
  T_{\text{loop}}/ \delta \mc T_{\text{loop}}^{(0)} \nonumber\\  
  &= \langle \wbar {\mc M_\varphi}({\bf B}) \rangle_L \equiv
  \frac{1}{L_{\text{esc}}} \int_0^{\infty} \d L\: \e^{- L/L_{\text{esc}}}\:
  \wbar {\mc M_\varphi} (L; \mathbf B) \; .
\label{qucor}
}
Note that current conservation, i.e., $\delta \mc R_{\text{diag}} +
\delta \mc R_{\text{loop}} = - \delta \mc T_{\text{loop}}$, is fulfilled in the
semiclassical limit $N, N^\prime \gg 1$. In the absence of SO
interaction, we have $\wbar {\mc M_\varphi} (L; \mathbf B) = (2s+1)\,
\exp(-\wtilde B^2 L/L_b)$, where $\wtilde B = 2 \sqrt 2 \pi B_z  \mt A_0/
\Phi_0$ and $\mt A_0$ is the typical effective area enclosed,
and the usual Lorentzian $\wtilde B$-profile \cite{bara,rich}
is recovered by Eq.\ \eq{qucor}.

In the case of SO interaction, the  quantum corrections \eq{qucor} depend on
the modulation factor $\wbar {\mc M_\varphi} (L; \mathbf B)$, which
characterizes the average spin evolution of a trajectory ensemble  and can be
easily determined from {\em classical} numerical simulations. Without $\bf
B$-field, $\wbar {\mc M} (L) \equiv \wbar {\mc M_\varphi} (L; 0)$ changes from
$\wbar {\mc M} (0) \!= \! 2s \! + \! 1$ to the asymptotic value $\wbar {\mc M}
(\infty) = (-1)^{2s}$ \cite{zait2}; i.e., for $s \!=\! 1/2$ an initial
polarization ($ {\mc M}(0) \!=\! 2$) becomes completely randomized ($ {\mc
M}(\infty)\!=\! -1$) owing to SO interaction, if the particle motion is
irregular (see below). Thus, if the particle quickly leaves the cavity (large
openings, small $L_{\text{esc}}$) or the SO interaction is too weak, there is
not enough time for the modulation factor to deviate from $2s + 1$, giving rise
to standard WL. In the opposite limit (large $L_{\text{esc}}$ or relatively
strong SO coupling) $\wbar {\mc M} (L)$ quickly reaches its asymptotic value
and, in view of \eq{qucor},  $\langle \wbar {\mc M_\varphi(0)} \rangle_L
\simeq $  $(-1)^{2s}$. Hence for half-integer spin the  conductance correction
becomes positive due to SO interaction. This phenomenon of weak
antilocalization does not exist for integer spin. For $\mathbf B \neq 0$ we
find $\wbar {\mc M_\varphi} (\infty; \mathbf B) = 0$ \cite{zait2}: Both a
magnetic flux (destroying constructive interference of the orbital phases) and
the Zeeman interaction (affecting the spin phases) inhibit AL.

For a quantitative treatment we must specify the form of the SO interaction. In
the following numerical analysis we consider the spin  $s \!=\! 1/2$-case for
different quantum dot geometries and Rashba SO coupling \cite{bych}, relevant
for 2D semiconductor heterostructures. It is described by an effective
magnetic field $\mathbf C \!=\! (2\, \alpha_R\, m_e / \hbar^2)\, \mathbf v
\!\times\! \hat{\mathbf z}$, where $\alpha_R$ is the Rashba constant, $m_e$ is
the effective mass, and $\mathbf v$ the (Fermi) velocity.  
%\footnote{$\alpha_R
%/ \hbar^2$ is kept fixed in the formal semiclassical limit $\hbar\! \to \!
%0$.}.  
In a billiard with fixed kinetic energy, $\mathbf C$ is constant by
magnitude and its direction changes only at the boundary. It is convenient to
characterize the  SO interaction strength by the mean spin-precession angle per
bounce, $\theta_R = 2 \pi L_b/ L_R$, where $L_R = 2 \pi |\mathbf v|/C$ is the
Rashba length. 

\begin{figure}[tbp]
  \includegraphics[width=.42 \textwidth, angle=0]{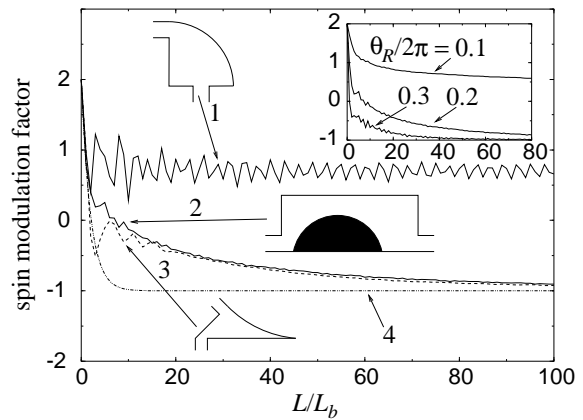}
  \caption{ Average modulation factor $\wbar {\mc M} (L)$ as a function
  of orbit length $L$ in units of bounce length $L_b$ for the quarter-circle
  billiard (curve 1), the desymmetrized Sinai billiard (curve 2), the
  desymmetrized diamond billiard (curve 3), and for an unbounded
  diffusive system with mean-free path equal to $L_b$
   (analytical curve 4). The
  relative strength of spin-orbit interaction is $\theta_R/2 \pi = 0.2$.
  \emph{Inset:} $\wbar {\mc M} (L)$ for the desymmetrized Sinai billiard at
  different values of $\theta_R$.
  \label{modfact}}
\end{figure}

In the inset of Fig.\ \ref{modfact} we plot the modulation factor $\wbar  {\mc
M} (L)$ for three SO strengths for a chaotic, desymmetrized Sinai (DS) billiard
(Fig.\ \ref{modfact}, geometry 2),  i.e., a prototype of a geometry with
hyperbolic classical dynamics. The average was performed over an ensemble of
$10^5$  (non-closed) trajectories (in the closed system) with random initial
velocity directions and positions  at the boundary. As  $\theta_R$ increases,
$\wbar {\mc M} (L)$ reaches its asymptotic value $-1$ faster.  In the main
panel of Fig.\ \ref{modfact} we compare, for fixed $\theta_R/2 \pi = 0.2$,
$\wbar {\mc M} (L)$ for four systems representing three different types of
orbital motion:

{\em Chaotic systems:} 
Two representative geometries, the DS billiard (curve 2)  and the desymmetrized
diamond (DD) billiard \cite{muel} (curve 3),  show up to deviations at small
lengths  nearly the same decay  behavior of $\wbar {\mc M} (L)$, indicating
\emph{universality} features for chaotic dynamics. 

{\em Integrable systems:} 
Although Eq.\ \eq{qucor} is valid only for chaotic cavities, the average
modulation factor $\wbar {\mc M} (L)$ is well defined for other types of
motion. Remarkably, we find that for the integrable quarter-circle (QC)
billiard (curve 1) $\wbar {\mc M} (L)$  oscillates  (with frequency independent
of $\theta_R$) around a constant saturation value well above $-1$. A systematic analysis shows \cite{zait2}  that the saturation
value in the integrable case is system-dependent and decreases down to $-1$,  indicating
spin relaxation, with increasing $\theta_R$. 

{\em Diffusive  systems:}
Unbounded diffusive motion (curve 4) exhibits fast exponential relaxation;
i.e., $\wbar {\mc M} (L) \simeq 3 \exp[- (\theta_R^{\, 2}/3) (L/L_b)]\! - \! 1$
\cite{zait2}, where $L_b$ is identified with the scattering mean-free path 
(cf.\ Eq.\ (10.12) of Ref.\ \cite{chak}). 

Note that the curves 1-4 almost coincide for $L \lesssim L_b$, because up to
the first scattering event the particle moves along a straight line, and
different types of dynamics cannot be distinguished. On larger length scales,
we find \emph{significant qualitative and quantitative differences in the spin
evolution in  chaotic, integrable, and diffusive systems.}  In particular, the
relaxation is strongly suppressed for a confined, even chaotic,  motion as
compared to an unbounded diffusive motion with the same $\theta_R$. This result
is supported by the following argument: In the limit $\theta_R \ll 1$ the spin
movements on the Bloch sphere ``mimic'' the orbital motion to order
$\theta_R^{\, 2}$; i.e., they are bounded for a spatially confined system.  If
higher-order corrections were neglected, the spin relaxation would saturate at
$L \sim L_b$. The further decrease of $\wbar {\mc M} (L)$, of order $(L/L_b)\,
\theta_R^{\, 4}$, is due to a Berry phase acquired by the spin wave function
\cite{alei,crem}. Its effect is similar to that of the AB phase. Hence, in a
chaotic system without Zeeman interaction one finds \cite{zait2} 
\eqn{
  \wbar {\mc M_\varphi} (L; \mathbf B) \simeq \e^{-(\wtilde B + \wtilde
  \theta_R^{\, 2})^2 L/L_b} + \e^{-(\wtilde B - \wtilde \theta_R^{\, 2})^2
  L/L_b},
  \label{mfweak}
} 
with $\wtilde \theta_R^{\, 2} = (\mt A_0/L_b^{\, 2})\, \theta_R^{\, 2} / \sqrt
2$. The further relaxation is due to terms of order $(L/L_b)\, \theta_R^{\, 6}$
\cite{alei,crem}. It eventually renders $\wbar {\mc M} (L)$ negative and
causes AL. For stronger interaction, $\theta_R \sim 1$, the three mechanisms
(initial relaxation, Berry phase, and further relaxation) work
simultaneously and cannot be separated (e.g., curves 2 and 3 in Fig.\
\ref{modfact}). 

\begin{figure}[tb]
  \includegraphics[width=.43 \textwidth, angle=0]{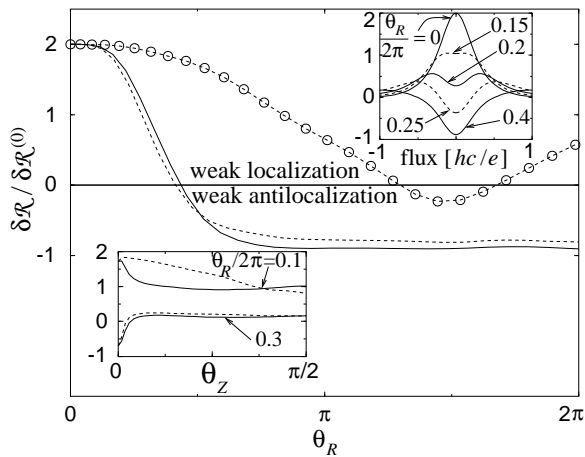}
  \caption{Relative quantum correction to the reflection $\delta \mc R/ \delta
  \mc R^{(0)}$ \emph{vs.}\ spin-orbit interaction $\theta_R$ for $\mathbf
  B\!=\! 0$ in the de\-sym\-metrized Sinai (solid), diamond (dashed), and
  quarter-circle (dashed with circles) billiards with $\mt P_c/(w + w^\prime) =
  90$. \emph{Lower left inset:} $\delta \mc R/ \delta \mc R^{(0)}$ \emph{vs.} 
  Zeeman interaction $\theta_Z$ for the de\-sym\-metrized Sinai billiard. The
  in-plane field is directed parallel (solid) and perpendicular (dashed) to the
  long side. \emph{Upper right inset:} $\delta \mc R/ \delta \mc R^{(0)}$
  \emph{vs.} perpendicular magnetic flux for the same billiard with $\theta_Z =
  0$.
  \label{antiloc}}
\end{figure}

Our numerical simulations show that in integrable systems both the spin
direction and the phase oscillate almost periodically during the orbital
motion.  Therefore,  after a short transient period, $\wbar {\mc M} (L)$ 
usually saturates. One exception we found is the circular billiard. Here, owing
to angular momentum conservation, all trajectories efficiently accumulate area,
and   $\wbar {\mc M} (L) \simeq 2 \sin (x)/x$ for $\theta_R \ll 1$,  where $x =
\theta_R^{\, 2} L r /2 L_b^{\, 2}$ and $r$ is the radius \cite{zait2}.

Figure \ref{antiloc} shows the relative quantum correction to reflection,
$\delta \mc R/ \delta \mc R^{(0)}$, as a function of $\theta_R$ for chaotic 
and integrable geometries (at $\mathbf B = 0$).  Positive (negative) values of
$\delta \mc R/ \delta \mc R^{(0)}$ indicate WL (AL). For chaotic systems
$\delta \mc R/ \delta \mc R^{(0)}\! \equiv \! (\delta \mc R_{\text{diag}} +
\delta \mc R_{\text{loop}})/(\delta \mc R_{\text{diag}}^{(0)} \!+\! \delta \mc
R_{\text{loop}}^{(0)}) \!=\! \langle \wbar {\mc M_\varphi} \rangle_L$ is given
by Eq.\ \eq{qucor}. For the numerical calculation of $\wbar {\mc M} (L)$ only
(backscattered) orbits starting and ending at one lead are considered (since
they are closed, the initial spin relaxation is reduced compared to non-closed
paths). The chaotic DS (solid curve) and DD (dashed curve) billiards show a
very similar WL-AL transition with increasing Rashba strength. The escape
length in units of $L_b$ is $\mt P_c/ (w + w^\prime)$, where $\mt P_c$ is the
perimeter of the cavity. Hence, given $L_R$, one can also conclude that AL is
absent in smaller quantum dots [for fixed $\mt P_c/ (w + w^\prime)$ or $w +
w^\prime$], as supported by experiment \cite{zumb}.  

The results for the integrable QC billiard (dashed curve with
circles) are based on a numerically obtained length distribution, which is no
longer exponential.  The transition to AL in the integrable billiard is 
much less pronounced and occurs at clearly higher $\theta_R$, compared to 
its chaotic counterparts owing to the slower spin relaxation. 
Hence there exists an extended regime of SO strengths, where one can switch
from WL to AL by tuning the classical dynamics from integrable to chaotic.

The Zeeman interaction, measured by a precession  angle $\theta_Z$ per bounce
(analogous to  $\theta_R$), suppresses AL (lower left inset). Note the
anisotropy in the (in-plane) field direction.  The  upper right inset depicts 
the magnetic-flux dependence.  The characteristic double-peak structure  
follows from Eqs.\ \eq{qucor} and \eq{mfweak}. 

The present semiclassical approach has a wider range of applicability,
including ballistic integrable systems and SO strengths up to 
$\theta_R \sim 1$, compared to RMT \cite{alei,crem}, 
which assumes  $\theta_R\! \ll\! 1$ in the ballistic regime (Eq.\  
(23) of \cite{crem}). Moreover,  the RMT results
contain free geometric parameters that have to be computed separately.

A corresponding analysis of ballistic conductance fluctuations \cite{bara} with
spin appears promising. We expect the \emph{shape} of its power spectrum to be
independent of  $\theta_R$ in an integrable system, but not in a chaotic
system.

We thank  M.\ Brack, A. V.\ Khaetskii, and M.\ Pletyukhov for stimulating
discussions and P.~Brouwer for a helpful clarification. The work has been
supported by the Deutsche For\-schungs\-gemein\-schaft (OZ and KR) and the  EU
Spintronics Research Training Network (DF).

\end{document}